# Charge transport in λ-DNA probed by conducting-AFM, and relationship with its structure.

Thomas Heim, Dominique Deresmes, and Dominique Vuillaume[a]
Institut d'Electronique, de Microélectronique et de Nanotechnologie - CNRS,
BP69, Avenue Poincaré, F-59652cedex, Villeneuve d'Ascq, France

87.14.Gg, 72.80.Le, 73.90.+f, 87.15.-v

*We studied the electrical conductivity of DNA samples as function of the number of DNA molecules. We showed that the insulating gap (no current at low voltage) increases from ~1-2 V for bundles and large ropes to ~4-7 V for few DNA molecules. From the distance dependent variation of the current, a unique hopping distance of ~3 nm is calculated (polaron-hopping model) independently of the number of DNA in the sample. The highly resistive behavior of the single DNA is correlated with its flattened conformation on the surface (reduced thickness, ~0.5-1.5 nm, compared to its nominal value, ~2 nm).*

Molecular electronics has attracted a growing interest owing to its envisioned possibilities to build high-density, low-cost, electronic circuitries. One of the challenging issues deals with the connection of a huge number of molecular-scale devices without the drawback of using traditional e-beam lithography for the patterning of electrical wires and contacts. Thus, the demonstration of a highly conducting molecular wire is crucial for future developments. In 1962, Eley and Spivey suggested that π-stacking in double-strand DNA (ds-DNA) could lead to an easy one dimensional charge transport.[1] Charge transfer (CT) through DNA molecules was widely studied for a large amount of DNA molecules in solution[2-6] because CT mechanisms have important implications in the damage and repair of this biological system. The conductivity was also studied at the "solid-state" in thin film of DNA-based compounds.[7] Recently, DNA molecules deposited on a solid substrate and connected between two electrodes were found highly conducting[8,9,10], insulating[11-14,15] or semiconducting.[16,17] These contradictions may come from differences in the base sequence, in the buffer and ambient conditions, in the structural organization of the DNA samples, in the number of DNA molecules in the sample (film, rope, single molecule), in the electrode/DNA coupling, etc…

In this letter, we report our experiments on CT in λ-DNA using conducting probe atomic force microscopy (C-AFM). We performed a systematic study of the distance-dependent behavior of the CT in DNA versus the size of the DNA samples: from DNA polymers, bundles and ropes to few single molecules. In that latter case, the CT behavior was correlated with the flattened conformation of the DNA molecule as observed by topographic TM(tapping mode)-AFM.

The λ-DNA was purchased from Roche-Biomedicals. The DNA molecules were dispersed in a TE buffer (Tris-HCl, EDTA) at pH~6.5. Before the deposition of NA, we chemically treated the $SiO_2$ surfaces with various silanizating agents (aminopropyltriethoxysilane, APTES, or octadecyltrichorosilane, OTS) or with spin coated polystyrene (PS). A drop (20 µL) of buffer containing DNA (at 250 ng/µL, i.e. 10 pM, otherwise specified) was deposited on these treated-surfaces and let to dry. Finally, the reference electrodes (~10 nm thick of gold or an organic conductor - pentacene) were vacuum ($10^{-8}$ Torr) evaporated through a shadow mask to contact the DNA molecules at one end (inset Fig. 1). We used a very sensitive ($10^{-15}$ A) home-made modified Digital Nanoscope III to do the C-AFM measurements. We topographically imaged the deposited DNA molecules by standard TM-AFM (tapping mode AFM). We recorded the I-V curves at a fixed position on the DNA molecules by appling the tip on the DNA with scanning parameters (x- and y-scans) fixed at zero and at a loading force of 10-30 nN.[18] All measurements were taken at room temperature in ambient air at a relative humidy (RH) of ~50%. It was reported that decreasing RH increases the resistivity of DNA.[19,20] A few of our C-AFM measurements (not reported here) taken under a dry nitrogen flux (RH<20%) confirmed this behavior. Thus, the data reported in this letter concern the DNA molecules with its hydration layer and counterions.

---

[a] Corresponding author: vuillaume@isen.iemn.univ-lille1.fr

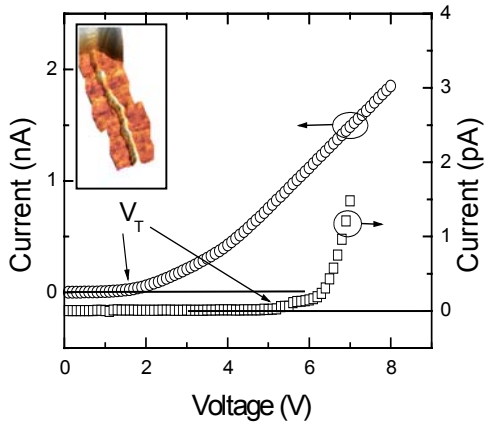

**Figure 1 :** *Typical current-voltage (I-V) curves: (○) For a large rope of ~2000 DNA on polystyrene substrate taken at the frontier between the rope and the bundle (see Fig. 1-a); (□) For a rope of about 500 DNA on polystyrene substrate and at a distance of ~1 µm. Inset: TM-AFM images of a very small ropes (~5 DNA) on amino-terminated surface. Deposition : 10 pM DNA, 10 mM TE. Image size : 500x500nm.*

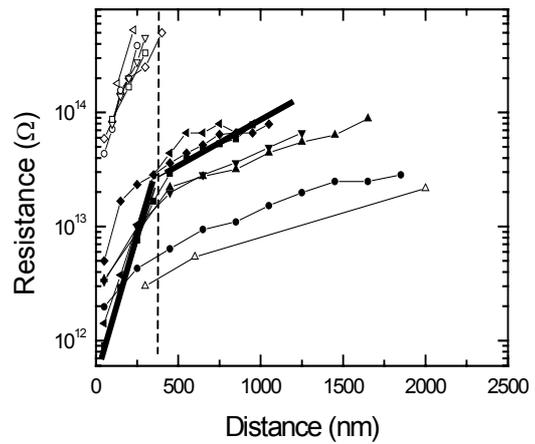

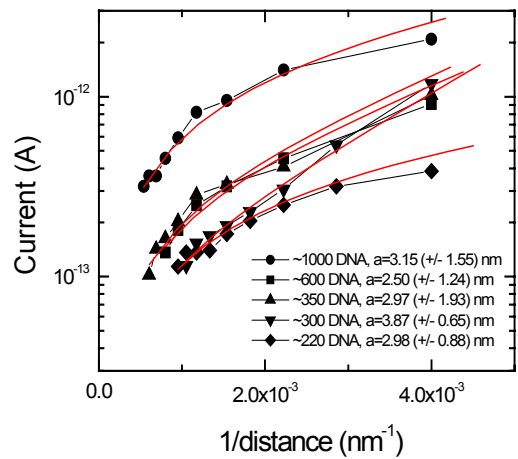

**Figure 2 :** *(a) Log-Lin plot of the resistance (measured from I-V curves above $V_T$) versus distance for a large number of samples with various sizes of the DNA sample: (●) ~1000, (△)~1000, (▼) ~600, (▲) ~350, (■) ~300, (◆) ~220, (○) ~150, (□) ~120, (◇) ~10, (▽) ~13 and (◁) ~5 DNA molecules. Closed symbols correspond to DNA on PS-treated surfaces, open symbols correspond to DNA on APTES-treated surface. The bold solid lines show exponential fits on the data (■) with two regimes. The typical fitted β values are : $\beta_{<300} = 1.01 \times 10^{-2}$ nm$^{-1}$ for d<~300 nm and $\beta_{>300} = 1.85 \times 10^{-3}$ nm$^{-1}$ for d>~300 nm. (b) Fits of the current-distance curves by the hopping model (eq. (1)): points are the experimental data, lines are the best fits with the hopping distance as the unique fit parameter. Other parameters in eq. (1) are : V=8V and T=300K. The fitted parameters are given on the figure.*

Figure 1 shows typical I-V curves along with the estimated number, N, of involved DNA molecules. A "blank" experiment with the C-AFM tip directly on the chemically treated-$SiO_2$ substrate near the DNA under test showed no measurable current (<$10^{-15}$ A). For the large samples (big ropes and networks, N>~1000), the I-V curves exhibited a plateau at null current (insulating gap) and a threshold $V_T$ of ~1-2 V at which the current increases. For the small ones (N<~1000), $V_T$ increases up to the range of ~4-7 V. A resistance is deduced from the first derivative of the I-V curves around a given bias (above $V_T$, typically at 7-8V). All measured DNA samples were highly resistive (Fig. 2), from R~$10^9$ Ω for bundles and very large ropes to $10^{15}$ Ω for few DNA molecules. Cai and coworkers measured by C-AFM resistances with the same order of magnitude ($10^9$-$10^{13}$).[13, 14] This also confirms results reported by de Pablo et al.[11] and Storm et al.[12] who showed R>$10^{12}$ Ω for distance larger than few tens of nanometers. We determined the resistivity per DNA molecule, $\rho_{DNA}$, using the estimated section area ($A=NA_{DNA}$ where $A_{DNA}$ is the nominal section of a single DNA molecule ~3nm$^2$) of the ropes and $\rho_{DNA}=A\ \partial R/\partial d$ (most of our R-d behaviors may be linearized over an enough small d range). We found that $\rho_{DNA}$ is more or less constant, $\rho_{DNA}$~5x$10^6$ Ω.cm, irrespective of the size of the measured DNA samples. This value is in agreement with De Pablo et al.[11] ($\rho_{DNA}$>$10^6$ Ω.cm), Storm et al. ($\rho_{DNA}$>~$10^5$ Ω.cm),[12] Okahata et al. ($\rho_{DNA}$~$10^5$ Ω.cm).[7]

To explain the distance-dependent behavior, an usual model (superexchange mechanism)[5, 21] is to fit the current according to an exponential attenuation, $I=I_0 e^{-\beta d}$, or $R=R_0 e^{\beta d}$, with a decay rate β. We distinguished two regimes (Fig. 2-a). For d<~300 nm, the fits give $\beta_{<300}$~$10^{-2}$-$10^{-1}$ nm$^{-1}$, while a smaller value is found for d>~300 nm, $\beta_{>300}$~$10^{-3}$-$10^{-4}$ nm$^{-1}$. Although smaller than the values in solution (1-15 nm$^{-1}$)[4, 5], $\beta_{<300}$ is in agreement with other experiments on solid surface.[13] The discreapency between "solution" and "solid-state" experiments may come from the difference in the conformation and surrounding environment of the DNA molecules (e.g. stretching, deformation…). At larger scale length, the very low β value ($10^{-4}$ nm$^{-1}$) seems unphysical (too low energy barrier) and we can discard this model. For d>300 nm, we observed the best agreement for the largest number of our experiments with the polaron-hopping model.[5, 11, 22-24] According to this model, the voltage dependence of the current is given by

$$I \propto \sinh(eaV/2kTd) \quad (1)$$

where a is the hopping distance, e the electron charge, V the applied bias (here above $V_T$), d the distance between the electrodes, T the temperature and k the Boltzman constant. Figure 2-b shows typical fits of our data with equation (1). We found a hopping distance of ~3-4 nm independently of the number of DNA in the ropes. This sample size independent parameter proves the good consistency of the model with the experiments. This value is a little bit larger than the one measured on poly(G)-poly(C) DNA (~2.5nm).[17] Since CT in DNA is more efficiently mediated through guanine base,[23-28] it is not surprising to have a larger hopping distance in λ-DNA (where the G bases are randomly distributed) than in poly(G)-poly(C). Moreover, this value (3 nm is about 8-9 base pairs), larger than the average G-G distance in λ-DNA (0.5-0.7 nm based on an average GC base pair content of 50-70%), may also account for intermolecular hopping in these entangled DNA samples.

The threshold voltage $V_T$ may be related to the difference between the Fermi energy of the electrodes and the molecular orbitals of the DNA ropes. The $V_T$ values is related to the average energy barrier that carriers have to overcome to be injected in the DNA molecules. The decrease of $V_T$ when increasing the number of DNA molecules in the sample suggests that the DNA molecular orbitals are closer to the electrode Fermi energy in large ropes and

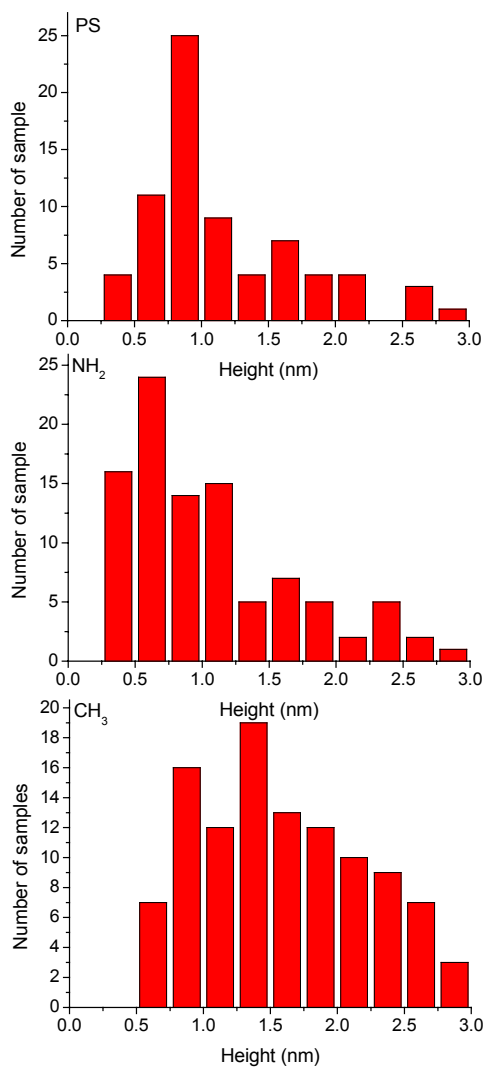

**Figure 3:** *Histograms of the DNA height measured by TM-AFM for single, isolated, molecules (or at least few molecules) deposited on various treated surfaces (PS, APTES - i.e. NH$_2$ terminated and OTS – i.e. CH$_3$ terminated).*

networks than in a single DNA. This feature may results from the well known reduction of the molecular orbital gap when organic systems move from single molecules to molecular materials.[29, 30]

Finally, a TM-AFM study on all samples with single (or a few) DNA molecules (whatever is the treated surface) showed that the height of the DNA molecule is smaller than the expected crystallographic value of ~2 nm. For the PS, APTES and OTS treated surfaces, the average heights were 1.17, 1.06 and 1.58 nm, respectively (Fig. 3), with a maximum of samples between 0.5 and 1 nm for PS and APTES surfaces and between 1 and 1.5 nm for OTS surfaces. This implies that the DNA molecules are distorted, flattened, when deposited onto the surface, and this feature could be responsible for the high resistivity reported here. Kasumov and Klinov[31] recently proposed the same conclusion. They found that a DNA molecule deposited on a pentylamine-treated mica surface has a height of about 2 nm and that it is more conducting than DNA on an untreated surface which has a height of ~1 nm. The second important geometric factor is the degree of stretching. From a fluorescence microscope study,[32] we found that only the DNA molecules on the amine-terminated surface have the nominal length of 16 μm, while DNA molecules are overstretched by an average factor of 1.7-1.75 on both PS and OTS surfaces, in agreement with previous report.[33] Thus the large consensus in the literature, that DNA is highly resistive, seems to be related with distorted nature of the DNA deposited on solid substrate and further study with a other treated surface, avoiding any distortion of the DNA, is mandatory to close the debate.

**Acknowledgements.** We thank fruitful discussions with H. Bouchiat (LPS-Orsay), V. Croquette (ENS-Paris).